\documentclass[12pt, hyperref]{article}

\RequirePackage{graphicx}
\RequirePackage{slashed} 
\RequirePackage{epsfig}

\def\({\left(}
\def\){\right)}
\def\[{\left[}
\def\]{\right]}
\def\a{\alpha}

\def\ep{\epsilon}
\def\g{\gamma}

\def\o{\omega}

\def\s{\sigma}
\def\bp{{\mathbf p}}
\def\bq{{\mathbf q}}
\def\G{{\Gamma}}

\def\be{\begin{equation}}
\def\e{\end{equation}}
\def\ee{\end{equation}}

\def\tr{{\rm tr}}

\title{Faraday rotation in graphene}

\author{I. V. Fialkovsky\\
    Instituto de F\'sica, Universidade de S\~ao Paulo,\\
     S\~ao Paulo, Brazil, e-mail:
    \emph{ifialk@gmail.com}\label{addr1}
          \and
        D. V. Vassilevich\\
                  CMCC, Universidade Federal do ABC, Santo Andr\'e,\\ SP, Brazil\label{addr2}\\
          Department of Physics, St.Petersburg State University,\\ St.Petersburg,
           Russia, e-mail: \emph{dvassil@gmail.com}\label{addr3}
}

\date{Received: date / Accepted: date}

\begin{document}

\maketitle
\begin{abstract}
We study magneto--op\-tical properties of monolayer graphene by means of quantum
field theory methods in the framework of the Dirac model. We reveal a good agreement between the Dirac model and a recent experiment on giant Faraday rotation in cyclotron resonance~\cite{giant_F}. We also predict other regimes when the effects are well pronounced. The general dependence of the Faraday rotation and absorption on various parameters of samples is revealed both for suspended and epitaxial graphene.
\end{abstract}

\section{Introduction}

Graphene, which is a one-atom thick layer of carbon atoms, does not need any lengthy introduction nowadays. Ever since its experimental discovery, graphene is in the spotlight of applied research in condensed matter physics~\cite{gra-rev1,gra-rev2,gra-rev3,gra-rev-11}.
The dynamics of quasiparticles in graphene is governed by the Dirac model
\cite{CastroNeto:2009zz}, i.e. they satisfy the Dirac equation where the speed of
light $c$ is replaced by the so-called Fermi velocity $v_F\simeq c/300$. Due to this
quasi-relativistic nature of electronic excitations, much of the graphene
physics is in fact described by Quantum Field Theory (QFT), see \cite{VKGrev,FVrev}
for recent reviews.

One the main obstacles to further industrialization of graphene is the gapless spectrum of  quasi--excitations, and thus most of the recent research was devoted to investigation of the electron transport in graphene sheets and possible ways of generating a controllable mass (energy) gap, see discussion in~\cite{gra-rev-11}. On the other hand, the optical properties of monolayer graphene from the beginning proved themselves extraordinary.
The very first measurement~\cite{abs1} of optical absorption in graphene revealed a surprisingly huge effect --- monoatomic layers absorbed about $2.3$\% of the incident light, exactly as predicted by the Dirac model, see \cite{abs2,ando, Fal, stauber}.

In this paper we investigate the optical properties of monolayer graphene samples in external magnetic field.
We focus here on the Faraday effect.
It may be thought of as an analog of the Hall effect considered  at ``non-zero frequencies'', and this similarity was used in Ref.\ \cite{VoMi} to conjecture that the
former should be common for Hall systems. In Ref.~\cite{ChS graphene} the Faraday rotation was related to possible
non-compensation of parity-odd parts of the polarization tensor between various generations
of fermions. Other related works on magneto--optical properties of monolayer graphene
include theoretical studies of the Kiev group, e.g. \cite{Gusynin_JPCond19_07,Gusynin_NJP11_09}, and numerical simulations in \cite{Aoki09}.

The quantity and quality of magneto-optical
experimental data is growing. In the case of multilayer graphene samples the first measurements of transmission spectra were reported in \cite{Potem_PRL97_06}, and the work was later continued in a series of papers, e.g. \cite{Potemski1,Potemski_multi}.
More recently, both the Faraday rotation and transmission in mono and multilayer epitaxial graphene were measured in \cite{giant_F}.
The Faraday rotation appeared to be unexpectedly large and was dubbed the ``giant Faraday effect".
The experimental setup of \cite{giant_F} permitted a detailed investigation of the so called cyclotron resonance regime.

To explore these effects we use a quantum field theory  approach to the physics of graphene.
Quasi--relativistic dispersion relation of the electronic excitations in graphene makes QFT a more adequate language to describe its properties as opposed to non--relativistic quantum mechanics. Application of the field--theoretical methods allow one to investigate such purely relativistic effects as the Klein paradox, or appearance of gauge fields in graphene. It also permits to investigate other ideas of QFT and quantum gravity in low energies bench--top experiments --- a situation hardly imaginable even ten years ago.

Following this approach, we start with the Dirac-Maxwell action, where the spinors are confined to the
graphene surface, while the electromagnetic field lives in the ambient $3+1$
dimensional space. Electromagnetic field is not quantized, and is given by a sum of
a constant magnetic field and a fluctuating one (photons). The magnetic field is taken
into account exactly, while in photons we keep the second order terms only. Therefore, on
the quantum field theory side we calculate a single one loop diagram (though a quite complicated
one, as the propagators contain the magnetic field, the chemical potential,
the mass, and a phenomenological parameter which describes impurities). This calculation
is similar to that of \cite{Gusynin_JPCond19_07,Gusynin_NJP11_09}, but is done differently
to facilitate a comparison to experimental data.

With some assumptions on the parameters (frequency, magnetic field, impurities, etc) one
can derive expression for conductivity of graphene and similar materials \cite{Tse2010,Flakov2011}
that are simpler than the ones given below.
We like to stress, however, that the conductivity that we calculate and use is an \emph{exact} result
of quantized Dirac model at the first order of the fine-structure constant $\alpha$ but without
any restrictions on the other parameters. The advantages of having such expressions (perhaps, complicated)
are obvious. Among them is the possibility of a cleaner judgement of validity of this or that approximation.
The price to pay is the use of the QFT machinery and terminology.

While the experiment of \cite{giant_F} can be fitted by the Drude model, the use of complete Dirac model makes it possible to relate the optical properties  of graphene to its microscopic parameters. It also permits to investigate and predict other regimes (besides the cyclotron resonance
one studied in \cite{giant_F}), where both the Faraday rotation and absorption are well pronounced.
In particular, we reveal quantization of the Faraday angle for clean samples, and predict pole--like peaks for relatively high frequencies.

Similar results were recently obtained in \cite{EqOM} using an approximate method of `equation of motion' for calculation of the conductivities of graphene.

We also like to mention the theoretical studies \cite{TI_Faraday1,TI_Faraday2} of the Faraday effect and related phenomena in topological insulators, which are closely related to graphene \cite{gra-rev-11}.

The paper is organized as following. In the next section we shortly review the basics of the Dirac model in graphene. In Section 3 we formulate its connection to optical properties of both epitaxial and suspended samples. In Section 4 we present the calculation of the polarization operator in magnetic field both for ideally clean and realistic samples, and finally in Section 5 we investigate the Faraday rotation and transmission of light performing the fit of existing experimental data, and predicting new regimes.

\section{The Dirac Model}
The Dirac model of the quasi--particles in graphene is based on the tight--binding model\footnote{
Magneto-optical properties of graphene can be analyzed \cite{Pe1,Pe2} starting directly with the
tight-binding model. This is technically more complicated than the analysis based on the Dirac model.
Both approaches coincide for the energies below $\sim 4.4{\rm eV}$ \cite{Pe2}.}.
An interested reader can find its detailed description in a number of reviews, e.g. \cite{CastroNeto:2009zz,Gusynin_IntJMP21_07}, while we briefly formulate the model itself.

Consider an (infinite) graphene sheet occupying the plane $x^3=0$ in a $3+1$
dimensional ambient space. The action for an electromagnetic field interacting
with quasiparticles propagating in the graphene surface reads
\be
    S= -\frac14 \int d^4x\, F^2_{\mu\nu}+
        \int d^3x
        \bar\psi \slashed{D} \psi
    \label{action}
\ee
with $ F_{\mu\nu}=\partial_\mu A_\nu-\partial_\nu A_\mu$,
\begin{equation}
     \slashed{D}\equiv \slashed{D}(A) = i\tilde\gamma^j(\partial_j+ieA_j) -\Delta\,.
\end{equation}
Greek letters denote the coordinates of the ambient space, $\mu,\nu=0,1,2,3$.
Latin letters $j,k=0,1,2$ correspond to coordinates on the graphene plane.
Since there are $N=4$ species
of fermions in graphene, the gamma matrices are in fact $8\times 8$, being a direct
sum of four $2\times 2$ representations (with two copies of each of the two inequivalent ones).
The matrices $\tilde\gamma^j$ are rescaled, $\tilde\gamma^0=\gamma^0$, $\tilde\gamma^{1,2}=v_F\gamma^{1,2}$, as compared to ordinary gamma-matrices
 $\gamma_0^2=-(\gamma^{1,2})^2=1$. $v_F\simeq 1/(300)$ is the Fermi velocity.
 We use the units where $\hbar=c=1$. $\Delta$ is the mass gap.
 The Maxwell action is normalized in such a way that
\be
    e^2\equiv 4\pi\alpha =\frac{4\pi}{137}.
    \label{e}
\ee

By averaging over the fermions in (\ref{action}) one arrives at an effective
action for the electromagnetic field, so that the full action becomes
\be
   S_A=  -\frac{1}4 \int d^4x\, F^2_{\mu\nu} + S_{\rm eff}(A)
   \label{Sa}
\ee
where
\begin{equation}
    S_{\rm eff}(A)\equiv-i \ln\det (\slashed{D}(A)).
    \label{seffdet}
\end{equation}
Since the action (\ref{action}) is quadratic in $\psi$ the expression (\ref{Sa}) is
exact.

We shall be interested in the scattering of light by a graphene sheet placed in a
strong external electromagnetic field. Therefore, it is convenient to split the
electromagnetic potential into a background and fluctuating parts,
\begin{equation}
A_\mu \to A_\mu+ A_\mu^{\rm b}.
\end{equation}
In the case of the Faraday effect the background part $A_\mu^{\rm b}$ describes a constant magnetic field $\bf B$ perpendicular to the graphene plate, while $A_\mu$ is the photons. The background part
has to be taken into account exactly. On the other hand, for our purposes it is enough
to expand $S_{\rm eff}$ up to the quadratic order in the radiation field $A_\mu$,
which is given by the one-loop diagram
\begin{eqnarray}
S_{\rm eff}(A)&=& A \ \raisebox{-3.75mm}
    {\psfig{figure=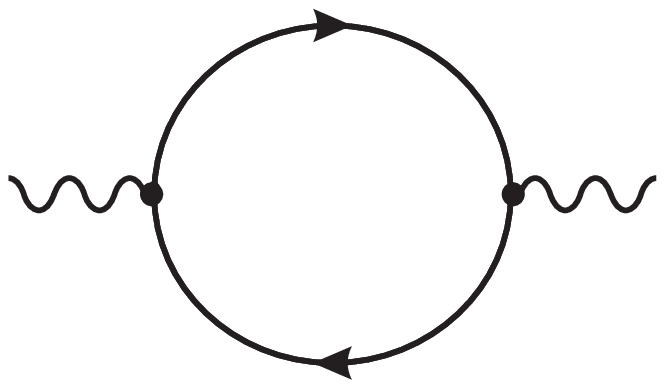,height=.4in}}
    \  A \nonumber\\
    &=& \frac 12 \int
    \frac{d^3p}{(2\pi)^3} A_j(-p) \Pi^{jk}(p;A^{\rm b})A_k(p),\label{Seff}
\end{eqnarray}
where the polarization operator $\Pi_{jk}$
\begin{eqnarray}
 && \Pi^{jk}(p) \nonumber\\
 &&=i e^2 \int \frac{dq_0 d^2 \bq}{(2\pi)^3}
        \tr\[
        \mathcal{S}(q_0,\bq)\tilde\gamma^j
        \mathcal{S}(q_0-p_0,\bq-\bp)\tilde\gamma^k\]
  \label{Pi_g}
\end{eqnarray}
depends on $A^{\rm b}$ through the fermion propagator $\mathcal{S}=(\slashed{D}(A^{\rm b}))^{-1}$. Here and below with bold
Latin letters we denote spatial in-plane vectors, $\bp=(p_1,p_2)$.

The propagation of electromagnetic waves in the whole space
can be described now by modified Maxwell equations with a delta-function interaction
corresponding to (\ref{Seff})
\be
    \partial_\mu F^{\mu \nu}+\delta(x^3)\Pi^{\nu \rho} A_\rho=0\,,
    \label{maxw}
\ee
where we  set $\Pi^{3\mu}=\Pi^{\mu3}=0$, $\mu =0,1,2,3$. These equations describe
free propagation of light away of the graphene subject to the matching
conditions
\be
\begin{array}{l}
  A_\mu\big|_{{x^3}=+0}=A_\mu\big|_{{x^3}=-0} \\
  (\partial_{3} A_\mu)_{{x^3}=+0}-(\partial_{3} A_\mu)_{{x^3}=-0}
    =\Pi_\mu^{\ \ \nu} A_\nu\big|_{{x^3}=0}
    \label{mc}
\end{array}
\ee
on its surface.
The same delta-function term in the Maxwell equations can also be interpreted as an in-plane current. Then the ac conductivity, that is a proportionality coefficient between the current ${\bf j}$ and electric field ${\bf E}$, ${\bf j}_a=\sigma_{ab}{\bf E}_b$, is expressed as
$$
\sigma_{ab}= \frac{\Pi_{ab}(\o)}{i \o}
$$
where $a,b=1,2$. This relation, where the polarization operator is considered independent of the spatial in-plane momenta, i.e. describing the normal incidence case, can be inverted
\begin{eqnarray}
 &&   \Pi^{jk}(\o,\bp=0)=\nonumber\\
 &&\(
                \begin{array}{cc}
                  \pi^0(\o) & \textbf{0} \\
                  \textbf{0} & i\o\(\delta^{ab} \sigma_{xx}(\o)
                    +\epsilon^{ab} \sigma_{xy}(\o)\) \\
                \end{array}
              \).
    \label{Pi_sigma}
\end{eqnarray}
Here $\delta^{ab}$ and $\epsilon^{ab}$ are two-dimensional Kronecker and Levi-Civita
symbols, respectively. By calculating explicitly the corresponding components of (\ref{Pi_g}) we shall find both diagonal and Hall ac conductivities, see section \ref{sec-pol}. The temporal component, $\pi^0(\o)$, will be of no importance in our further considerations.

\section{Optical properties of graphene}

Let us consider a beam of light linearly polarized along the $x\equiv x^1$ axis passing normally through a graphene layer on a substrate of a finite thickness $d$. In a magnetic field the Hall conductivity of graphene $\sigma_{xy}$ is non-zero. Therefore both $x$-- and $y$--polarizations will be present in the transmitted light.
The intensity of light measured
after passing graphene, substrate \emph{and} a polarizer rotated by angle $\phi$ in respect to the $x$--axis is given by \cite{pvt}
\begin{eqnarray}
    I_\phi&=&\frac{|t_{xx}|^2+|t_{xy}|^2}2+
\frac{|t_{xx}|^2-|t_{xy}|^2}2\cos2\phi\nonumber\\
    &&-{\rm Re}(t_{xx}^{\phantom{*}}t_{xy}^*)\sin2\phi,
    \label{Intens}
\end{eqnarray}
where $t_{xx,xy}$ are the transmission coefficient of $x$, $y$--polari\-za\-tions, respectively.

The angle $\theta$ where (\ref{Intens}) attains its maximum as a function of $\phi$ gives the angle of polarization rotation
\begin{equation}
    \theta=-\frac12\arctan
\frac{2{\rm Re }(t_{xx}t_{xy}^*)}{|t_{xx}|^2-|t_{xy}|^2}\,.
\end{equation}
The total transmission is just a sum of the intensities in two polari\-za\-tions,
\begin{equation}
    \tilde T={|t_{xx}|^2+|t_{xy}|^2}.
\end{equation}

By solving (\ref{mc}) with $\Pi$ defined by (\ref{Pi_sigma}) along with a standard matching condition at the dielectric--vacuum interface at
$z\equiv x^3=d$, we obtain
\begin{eqnarray}
&&t_{xx}=-4n_s e^{id(n_s-1)\o}\frac{A_{x}}{A_{x}^2+A_{y}^2}\\
&&t_{xy}= 4n_se^{id(n_s-1)\o}\frac{A_{y}}{A_{x}^2+A_{y}^2}
\end{eqnarray}
where
\begin{eqnarray}
&&A_{x}=(n_s-1)(n_s-1- \s_{xx})e^{2id\o n_s}-( n_s+1)( n_s+1+\s_{xx})\nonumber\\
&&A_{y}=\s_{xy}\(n_s+1+(n_s-1)e^{2id\o n_s}\)\,,
\end{eqnarray}
%
and $n_s$ is the refractive index of (non--absorbing) substrate.

Both polarization rotation angle and transmission oscillate rapidly with the frequency $\omega$
(or with the substrate width $d$).
This effect is called the Fabry--Perot oscillations.
However, in many cases the oscillations are smeared out due to low resolution of the measurements, or other sources of incoherence. The averaged intensity can be obtained by integrating (\ref{Intens}) in $d$ over the period of oscillations, $d_p=\frac{\pi}{ \o n_s}$.

Thus, we have
\begin{eqnarray}
&&\theta=-\frac12\arctan\frac{8 {\rm\ Re}(\s_{xx}\s_{xy}^*)+4 (n^2_s+3){\rm Re}\s_{xy}}
        {|2\s_{xx}+n_s^2+3|^2-4|\s_{xy}|^2-(n^2_s-1)^2} \nonumber\\
&&
\tilde T=\frac{8n_s^2}{|a_+|^2-|b_+|^2}+\frac{8n_s^2}{|a_-|^2-|b_-|^2}
    \label{TtSubs}
\end{eqnarray}
where $a_\pm=(n_s+1)(n_s+1+\s_{xx}\pm i \s_{xy})$, $b_\pm=(n_s-1)(n_s-1-(\s_{xx}\pm i \s_{xy}))$,  $\tilde T$ was first obtained in \cite{Potemski1}.
Without graphene, the transmission of the bare substrate is
\begin{equation}
T_0=\frac {2n_s}{1+n_s^2}.
\end{equation}
Later on we shall use the normalized transmission,
\begin{equation}
T=\tilde T/T_0\,,
    \label{relT}
\end{equation}
which characterizes better the properties of graphene itself.

To obtain the first order expressions in $\a$ one
keeps only the linear terms in the conductivity
\be
    T\simeq 1 - {2f_s {\rm Re}\sigma_{xx}}, \qquad
    \theta\simeq -f_s  {\rm Re}\sigma_{xy}
    \label{Tth_a}
\ee
where the coefficient $f_s$ reads
\begin{equation}
    f_s=\frac{n_s^2+3}{4 n_s^2+4}.
\end{equation}

To reproduce the case of suspended graphene one puts $n_s=1$ (i.e. $f_s=1/2$) to obtain\footnote{The reflection
and transmission coefficients for suspended graphene
can be read off from \cite{ChS graphene} after the identification
$\sigma_{xx}=i\alpha\Psi /\omega$, $\sigma_{xy}=\alpha\phi$.}
\begin{eqnarray}
&&    T=4\, \frac{ |\s_{xx}+2|^2+|\s_{xy}|^2}{| \s_{xx}^2+\s_{xy}^2 +4 \s_{xx}+4|^2},\nonumber\\
&&    \theta=-\frac12{\rm arg} \frac{2+\s_{xx}+i\s_{xy}}{2+\s_{xx}-i\s_{xy}}
    \label{Tth}
\end{eqnarray}
while in the linear order in the conductivity the expressions are
\be
    T\simeq 1 - {\rm Re}\sigma_{xx}, \qquad
    \theta\simeq -\frac 12  {\rm Re}\sigma_{xy}\,.\label{Ttsus}
\ee

\section{Calculation of the polarization operator}\label{sec-pol}
The polarization operator $\Pi_{jk}$ has been considered in a number of papers for systems characterized by various sets of parameters. In the zero-temperature case it was
first calculated in \cite{Appelquist:1986fd}.
At finite temperature calculations were done in \cite{DoreyMavratos92},
while some other cases were considered in, e.g.,  \cite{LiLiu10,Pyatkovskiy_JPCM21_09,Vozm94}.
In the presence of an external magnetic field extensive calculations of the polarization operator  were done by the Kiev group
and collaborators\cite{Gusynin_PRB66_02,Gusynin_PRB73_06,Gusynin_JPCond19_07,
Gusynin_NJP11_09,Pyatkovskiy_JPCM21_09,Pyatkovskiy_PRB83_11}. Here we shall only sketch the
derivation.

In a constant magnetic field $B$ perpendicular to the surface of graphene
the fermion Green function has the form \cite{Chodos:1990vv,Gusynin_PRB73_06}
\be
    \mathcal{S}(\o,\bp)=e^{-\bp^2/|eB|}
        \sum_{n=0}^\infty(-1)^n\frac{\mathcal{S}_n(\o+\mu,\bp)}{(\o+\mu)^2-M_n^2}\,,
        \label{prop}
\ee
where
\begin{eqnarray}
  &&  \mathcal{S}_n(\o,\bp)=2(\o\g^0+\Delta)\[P_- L_n\(\frac{2\bp^2}{|eB|}\)
  \right. \nonumber\\
  &&\ \ \left. -P_+ L_{n-1}\(\frac{2\bp^2}{|eB|}\)\]
        -4v_F\bp\cdot{\mathbf \gamma}L_{n-1}^{(1)}\(\frac{2\bp^2}{|eB|}\)\,.
\end{eqnarray}
Here $ P_\pm=(1\pm i \g^1\g^2 \epsilon_B)/2$, $\epsilon_B\equiv{\rm sign}(B)$,
the Landau levels are $M_n=\sqrt{2nv_F^2 |eB|+\Delta^2}$, and
 $L_n^{(\alpha)}$ are the associated Laguerre polynomials, $L_n\equiv L_n^{(0)}$,
 $L^{(\alpha)}_{-1}=0$.  $\mu$ denotes the Fermi energy shift (or the chemical potential in the QFT terminology).

Next, we need to introduce
a phenomenological parameter $\Gamma$ which describes the presence of impurities.
This can be done by means of the substitution
$$
\o \to \o + i \Gamma{\rm\ sgn}\o, \quad \G>0.
$$
everywhere in (\ref{prop}).
In the limit $\Gamma\to 0$ one recovers the usual Feynman propagator.
For $\Gamma\ne 0$ the propagator $\mathcal{S}$ is not an analytic function
of $\omega$ due to the presence of ${\rm sgn}\, \omega$.
Most generally,  $\Gamma$ can dependent on frequency, magnetic field, etc.
We will restrict ourselves to a constant $\Gamma$, which is
sufficient in a not too strong magnetic field \cite{G_cnst_1}, but see also \cite{G_cnst_2}.

Describing the disorder in such a simplified manner, we assume that the long--range impurities present in considered samples of graphene are sufficiently weak (if any). Otherwise the states near the Dirac points get localized, and deviations from the Dirac dispersion should be taken into account. The behavior of graphene in
the presence of strong long-range impurities is studied in detail in \cite{DisOrd}.

We are ready to substitute (\ref{prop}) in Eq.\ (\ref{Pi_g}).
Since we shall consider normal incidence wave only, $\bp=0$ in (\ref{Pi_g}),
the loop integral over spatial momenta $\bq$ can be easily performed
with the help of the orthogonality condition for the Laguerre polynomials
\begin{equation}
\int_0^\infty x^\alpha e^{-x}L_{n}^{(\alpha)}(x)L_{m}^{(\alpha)}(x) dx=
\frac{(n+\alpha)!}{n!} \delta_{n,m} \,.\label{LLL}
\end{equation}

Thus, we arrive at the following expression for the Hall conductivity
\be
\sigma_{xy}(p_0)=\frac{\epsilon_B  \a N L_b^2}{p_0}\sum_{n=0}^\infty G_n 
      \label{s_xy}
\ee
where $L_b^2=2v_F^2 |eB|$, $G_n\equiv \int_{-\infty}^\infty d\o \(f_{n,n+1}-f_{n+1,n}\)$ and $f_{nm}$ is given by
\begin{eqnarray}
   && f_{nm}=\frac{i}{2\pi}  [(\tilde\o+i\Gamma {\rm\ sgn}\o)(p_0-\tilde\o-i\Gamma {\rm\ sgn}(\o-p_0))+\Delta^2]\nonumber\\
   && \ \ \times
        ((\tilde\o+i\Gamma {\rm\ sgn}\o)^2-M_n^2)^{-1} \nonumber\\
  &&\ \  \times ((\tilde\o-p_0+i\Gamma {\rm\ sgn}(\o-p_0))^2-M_m^2)^{-1}
    \label{fmn}
\end{eqnarray}
Here $\tilde\o\equiv \o+\mu$.
For the diagonal conductivity,  we have
\be
\sigma_{xx}(p_0)= \frac{i \a N L_b^2}{p_0} \sum_{n=0}^\infty H_n 
      \label{s_xx}
\ee
where $H_n\equiv \int_{-\infty}^\infty d\o \(f_{n,n+1}+f_{n+1,n}\)$.

Note, that since for describing graphene the $\gamma$-matrices are taken in a reducible representation
consisting of equal number of inequivalent irreducible ones related to each other though the parity transformation,
we have, in particular, ${\rm tr}\, (\gamma^0\gamma^1\gamma^2)=0$.

\subsection{QHE in clean graphene}
Calculations that we present  here are rather standard, but differ in details
from other sources. As a consistency check we reproduce in this subsection the
Hall effect in graphene.

For a clean graphene, $\G\to+0$, the calculation of the frequency integral in (\ref{s_xy}) can
be performed with the help of the Cauchy theorem yielding
\be
\sigma_{xy}(p_0)
={-\epsilon_B \a N L_b^2}\,
        \frac{p_0^2-M_{n_0}^2-M_{n_0+1}^2+2\Delta^2}
            {(p_0^2+M_{n_0+1}^2-M_{n_0}^2)^2-4 p_0^2 M_{n_0+1}^2}
    \label{s_xy_0}
\ee
where $n_0$ is defined in such a way that
$M_{n_0}<\mu$, while $M_{n_0+1}>\mu$,
i.e.
\be
    n_0= \left\lfloor
    \frac{\mu^2-\Delta^2}{2v_F^2 |eB|}\right\rfloor
    \label{n_0}
\ee
here $\left\lfloor x \right\rfloor$ denotes the integer part of $x$, defined so that $\left\lfloor x \right\rfloor =0$
for $x<0$. As expected, (\ref{s_xy_0}) coincides with the previous results and, for instance,
 with the $T\to 0$ limit of Eq.\ (12) in Ref.\ \cite{Gusynin_JPCond19_07}.

In the dc limit, $p_0\to0 $, one has
\be
    \sigma_{xy}^{dc}  =\lim_{p_0\to0}\sigma_{xy}
    = {\alpha\epsilon_B N}\(1+2 n_0\)\,. \label{Hall}
\ee
This again coincides with the results of other calculations of relativistic Hall conductivity
\cite{Gusynin_PRB73_06,TQHE1,TQHE2,BS,BGSS}. The law (\ref{Hall}), which is called
the anomalous quantum Hall effect, or the unconventional integer quantum Hall effect, was
checked on experiments \cite{QHE1,QHE2}. This is one of the most spectacular confirmations
of the Dirac model model of quasiparticles in graphene.
 Note, that this result holds for any value of the mass gap.

In the opposite limit, $p_0\to\infty$, (i.e., for the visible light), we get
\be
\sigma_{xy}\approx -\frac{ \epsilon_B \a N L_b^2}{p_0^2}.
\ee

\subsection{Impact of impurities}

In realistic samples impurities are always present,
so that one should keep $\Gamma$ positive. In this case, the fermion Green function, ${\cal S}$, is not an analytical function of
$\omega$, and we cannot apply the residue theorem for $\o$--integration in (\ref{s_xy}) and  (\ref{s_xx}). Still, either this integration or $n$--sum can be resolved explicitly there, and it is a matter of convenience which operation to perform first.
For analysis of the conductivity dependence on the frequency we find it more suitable to integrate first over $\o$.

After rather elementary but cumbersome algebra the integration in
the off-diagonal part gives
\begin{eqnarray}
   && G_n=\frac{i}{8\pi}\sum_{\ep,\ep'=\pm}\(\frac{g_1(\ep M_n,\ep'
 M_{n+1})}{p_0-(\ep M_n+\ep' M_{n+1})}\right.\nonumber\\
&&\quad \left.+        \frac{g_2(\ep M_n,\ep' M_{n+1})}{p_0+2i\G-(\ep M_n+\ep' M_{n+1})}\)
    - (\mu\to-\mu)\,,
    \label{Gn}
\\
&&g_1=\(1+\frac{\Delta^2}{M_{n+1}M_{n}}\)
\nonumber\\
&& \quad \times    \log\frac{(p_0+i\G+\mu-M_{n+1})(i\G+\mu+M_{n+1})}{(p_0+i\G+\mu-M_{n})(i\G+\mu+M_{n})} \,,
    \label{g_i}
\\
&&g_2=\(1+\frac{\Delta^2}{M_{n+1}M_{n}}\)\nonumber\\
&& \quad \times \log\frac{(p_0+i\G+\mu-M_{n})(i\G+\mu-M_{n+1})}{(p_0+i\G+\mu-M_{n+1})(i\G+\mu-M_{n})}\,.
\end{eqnarray}
The $\G\to0$ limit taken in the above expression restores (\ref{s_xy_0}).

In the diagonal part of conductivity,  we obtain a similar result
\begin{eqnarray}
&&H_n=\frac{i}{8\pi}\sum_{\ep,\ep'=\pm}\(
    \frac{h_1(\ep M_n,\ep' M_{n+1} )}
        {p_0-(\ep M_n+\ep' M_{n+1})} \right. \nonumber\\
&&\qquad  \left. +\frac{h_2(\ep M_n,\ep' M_{n+1})}{p_0+2i\G-(\ep M_n+\ep' M_{n+1})}\)
        +(\mu\to-\mu) \,,
        \label{Hn}\\
&&
h_1= \(1+\frac{\Delta^2}{M_{n+1}M_{n}}\)\nonumber\\
&&\quad \times    \log\frac{(i\G+\mu+M_{n})(i\G+\mu+M_{n+1})}
        {(p_0+i\G+\mu-M_{n})(p_0+i\G+\mu-M_{n+1})}
    \label{h_i}\,,\\
&&
h_2 = \(1+\frac{\Delta^2}{M_{n+1}M_{n}}\)\nonumber\\
&&\quad \times   \log\frac{(p_0+i\G+\mu-M_{n})(p_0+i\G+\mu-M_{n+1})}
        {(i\G+\mu-M_{n})(i\G+\mu-M_{n+1})}\,.\nonumber
\end{eqnarray}
It is clear
from (\ref{Gn}) and (\ref{Hn}), that  the off-diagonal part of the conductivity is odd in chemical potential, while the diagonal one is even.


\subsection{Renormalization}\label{sec-Ren}
The polarization operator (\ref{Pi_g}) is power--counting divergent in ultraviolet. It indeed shows up in the diagonal part (\ref{s_xx})
since
\begin{equation}
H_n \mathop\simeq_{n\to\infty} -\frac{1}{2 M_n}+O(n^{-3/2})\,.\label{Hass}
\end{equation}
and the infinite sum in $n$ does not converge. As we can see, the divergency is in the imaginary part of $\sigma_{xx}$, while in the $\a^1$ order we have $1-T\sim {\rm Re}\sigma_{xx}$ (\ref{Tth_a}). Thus, if one is interested in
the $\alpha^1$ level only the renormalization procedure is not required.
Generically it can be handled, for example, via the Pauli--Villars substraction at infinite mass.

In doing so, one considers a difference between two polarization operators (\ref{Pi_g}) taken at different masses $\Delta$, $\tilde\Delta$, where the latter one shall be taken to infinity after the loop momenta calculation. In (\ref{s_xx}) it results in substitution of $H_n(\Delta)$ by $H_n(\Delta)-H_n(\tilde\Delta)$. To calculate the asymptotics of such $\sigma_{xx}$ at $\tilde\Delta\to\infty$ one adds and subtracts (\ref{Hass}) from under the $n$--summation and then applies the Abel--Plana formula to express the sums via integrals, performs the expansion in large $\tilde\Delta$, and then takes the limit. The renormalized diagonal conductivity obtained in such a way reads
\begin{eqnarray}
 &&   \sigma_{xx}^R=\frac{\a i N L_b^2}{p_0}
        \(\sum_{n=0}^\infty \(H_n+\frac{1}{2 M_n}\)
            +\frac{\Delta}{L_b^2}
            -\frac1{4\Delta}
            -\frac{2\G}{\pi L_b^2}\right.
\nonumber\\
   &&\qquad   \left.   -\frac{i}2\int_0^\infty \frac{dx}{e^{2 \pi
 x}-1}\(\frac1{M(i x)}-\frac1{M(-i x)}\)
        \)\,,
\end{eqnarray}
where $M(z)=+\sqrt{ z L_b^2+\Delta^2}$. In Sect. \ref{cycl} we will use
this expression both for finite mass, and in the $\Delta\to0$ limit. The later one is consistent with the use of Pauli-Vilars regularization since
in $2+1$ dimensions there are no logarithmic divergencies unlike the $3+1$ dimensional case.

For vanishing $\mu$, $\Delta$ and $B$ the one-loop diagram in (\ref{Seff}) is finite, see \cite{Vozm94}. In our case this diagram is divergent since we have
considerably less symmetries. We would like to mention, that one has to be
careful with power-counting divergent diagrams that are finite for
symmetry reasons. Sometimes, such diagrams nevertheless require a
finite renormalization. For example, this effect gives rise to
to the parity anomaly \cite{Semenoff,Redlich84} in $2+1$ dimensions.

\section{Phenomenology}
The above formulas can be used now to investigate optical properties of realistic graphene samples, including those on a substrate. In particular, in the next subsection we show that the Dirac model nicely fits the results of the experimental measurements of magneto--optical properties of epitaxial graphene \cite{giant_F}.

Although the polarization rotation and absorption are very strong for a
one-atom thick material, these effects are still sufficiently small
to allow for a qualitative description by  the $\a^1$--order
of the perturbation theory. It is sufficient to use (\ref{Tth_a}) and (\ref{s_xy},\ref{s_xx}) in order to estimate the order of magnitude of the effect for transmission and rotation angle
\begin{eqnarray}
&&  T= 1 + 2f_s\frac{\a N L_b^2}{p_0}{\rm\ Im}\sum_{n=0}^{\infty} H_n\,,
\nonumber\\
&&   \theta= -f_s  \frac{\epsilon_B \alpha N L_b^2}{p_0} {\rm Re}\sum_{n=0}^\infty G_n\,.
    \label{T_th_wrk}
\end{eqnarray}
Still, for the correct numerical comparison with the results of the experiment one should use the exact formulas (\ref{TtSubs}) as discussed in the next subsection.

To distinguish between different physical regimes, we shall start by investigating the analytical behavior of both $T$ and $\theta$.
Contrary to the naive expectations, neither $G_n$ nor $H_n$ as functions of frequency $p_0$ has actually any poles in the complex plane.
%
%
%
%
%
Indeed, if you consider, for example,  $g_2(p_0,-M_n,M_{n+1})$ (\ref{g_i}) in the vicinity of $p_0=M_{n+1}-M_n-2i\G$ you find out that it vanishes there and thus cancels out the corresponding denominator. Similarly it behaves at all other points $\ep M_n+\ep' M_{n+1}$, etc.

On the other hand, in the clean graphene the Hall conductivity (\ref{s_xy_0}) (and thus the polarization rotation angle) has poles at
\be
    p_0=\pm(M_{n_0+1}\pm M_{n_0}).
    \label{poles}
\ee
One would naturally expect that the behavior of conductivity in realistic graphene samples with sufficiently small $\G$--s is somewhat similar to the clean case. Thus, in what follows we shall concentrate on these points and investigate the optical properties in their vicinity.

Two main physical regimes can  be distinguished already by considering (\ref{poles}). For a large chemical potential, $\mu\to\infty$, the system enters into the so called cyclotron resonance regime, which is characteristic to epitaxial graphene in magnetic field. The other limit, $\mu\to0$,  represents  suspended gra\-phene samples without gate voltage applied. The (positive) frequencies (\ref{poles}) in these cases are
\be
    p_0\ \mathop{\simeq}_{\mu\to\infty}\ \frac{L_b^2}{2\mu},\qquad
p_0\ \mathop{\simeq}_{\mu\to\infty}\ 2\mu,\qquad
        p_0\  \mathop{\simeq}_{\mu\to0}\   M_1.
        \label{w0}
\ee
In what follows we consider these two regimes in more details to correctly estimate the magnitude of the effects, and specify the position of their maxima.

\subsection{Epitaxial graphene, big chemical potential}\label{cycl}
The epitaxial graphene is characterized by a considerable Fermi energy shift due to the interaction with the atoms of the substrate. In known experimental devices it is of the order of tenths of eV, $\mu\equiv \epsilon_F\sim 0.1\div0.5$eV \cite{Potemski1,giant_F,Efermi1}. If a gate voltage is applied the chemical potential can be increased further more.
The other parameters of the system such as  the scattering rate or the distance between corresponding neighboring Landau levels (at moderate magnetic fields of the several Tesla) are of the order of $1\div10$meV, or smaller. Thus the chemical potential gives the highest scale, and the system is in the cyclotron resonance regime.

\begin{figure}
\centering
\includegraphics[width=8cm]{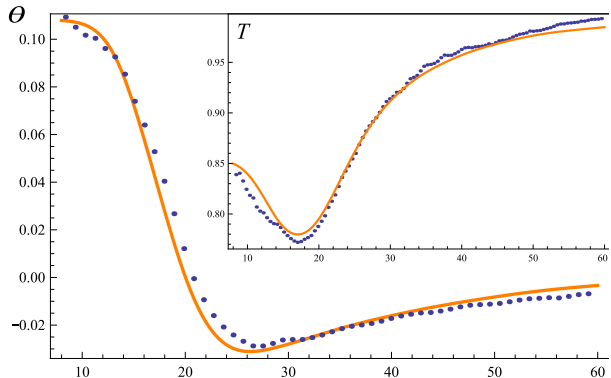}
\caption{Experimental data (dotted lines) and the best fit curves (solid lines, $\mu=289.2$meV, $\G=4.4$meV) for
the Faraday rotation and relative transmission (insert) at $\mathbf{B}=7$T as
functions of $p_0$[meV]. }
\label{fit7tesla}
\end{figure}

The experimental measurements of the Faraday rotation and intensity transmission in this regime were recently reported in \cite{giant_F}. In their set-up the authors used mono and multilayered graphene samples on a SiC substrate subject to a magnetic field varying from $0.5$T to $7$T. It is the aim of the current subsection to show that the Dirac model fits nicely the experimental curves for monolayer samples.
In what follows we focus on the Faraday rotation angle, but similar considerations can be given to the transmission as well.

Among the two points given in (\ref{w0}) for large chemical potential,
we shall consider the first one only, since the second one $p_0\simeq 2\mu$
corresponds to the frequencies that are too high. Qualitatively, the behavior
of the Faraday rotation angle at the second point is similar to
the one depicted on Fig.~\ref{on_p0} below.

Before performing the numerical fitting we note  that the main contribution to the infinite sum in (\ref{T_th_wrk}) comes from a small vicinity of $n=n_0(\mu)$, (\ref{n_0}). It follows from analytical investigation of ${\rm Re}G_n$, which also shows that for big enough $\mu$, roughly for $\mu > \frac{M_1^2}{2 \pi\G}$, each contribution to the sum has a maximum as a function of frequency at $p_0=0$
$$
    {\rm Re} \frac{\partial (G_n/p_0)}{\partial p_0}\Bigg|_{p_0=0}=0.
$$
The maximum value, ${\rm Re}G_n(p_0=0)$, as a function of $n$ for fixed chemical potential, is sharply peaked around $n=n_0$, the localization being governed by the difference of two slightly dislocated arctangents.

Thus, for comparing with observations we can replace the infinite summation in (\ref{s_xy},\ref{s_xx}) by a finite sum. For the given experimental data of \cite{giant_F}, it proved to be sufficient to choose the summation range for $n$
from $0$ to $n_0+30$. A rough estimate of the amplitude of the effect can be  given by the main contribution with $n=n_0$,
\begin{eqnarray}
&&\theta_{max}\equiv
    -f_s  \frac{\epsilon_B \alpha N L_b^2}{p_0}
{\rm Re}G_{n_0}\Bigg|_{p_0=0}\nonumber\\
 &&\qquad   \sim
    \frac{0.75 f_s \a L_b^2}{({M_{n_0+1}- M_{n_0}})^2}
    \sim \frac{3 f_s  \a \mu^2}{L_b^2},
\end{eqnarray}
which is in a reasonable agreement with direct numerical calculations (as well as with experimental results, provided we put $f_s\simeq0.31$ according to \cite{giant_F}).

\begin{figure}
\centering
\includegraphics[width=8cm]{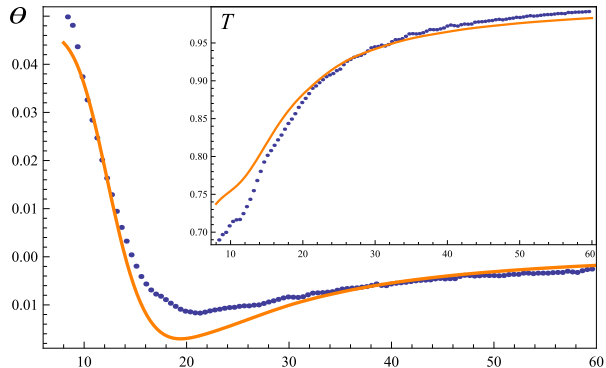}
\caption{The rotation and transmission (insert) theoretical predictions (solid lines) plotted against the frequency $p_0$[meV] at B$=2$T for $\mu=120$meV, $\G=3.9$meV, $\Delta=110$meV. Corresponding experimental curves are given in blue dotted lines.
}
\label{fit_alt}
\end{figure}

Despite that the approximate formulas (\ref{T_th_wrk})  give a correct order of magnitude estimate of the observed effect, to perform precise numerical fitting of the experimental data we should use the exact formulas (\ref{TtSubs}). The results of the fitting which was performed simultaneously for polarization rotation and relative transmission (as defined by (\ref{relT})), are summarized below:
\begin{itemize}
  \item
  The experimental curves for rotation and transmission are nicely fitted by the Dirac model at
    fields greater then $2$T.  The fit is exemplified on Fig. \ref{fit7tesla} where both experimental (dotted lines) and theoretical curves (solid ones) 
    are presented.
%

  \item The best fit value of the chemical potential  $\mu$ (i.e. the parameter which enters the Dirac
    model) is found for such fields to be approximately $280\div290$meV, somewhat lower then the Fermi energy shift measured at zero field, $\epsilon_F =340$meV. It does not show any specific dependence on the magnetic field.
  \item The fit turns significantly worse for magnetic field of $2$T and lower.
 However, it can be somewhat improved
    by greatly lowering the chemical potential (down to about $100\div120$meV)
and introducing a non--zero mass gap $\Delta$, see Fig. \ref{fit_alt}.

\end{itemize}

We do not perform a statistical analysis of the data and do not calculate
the accuracy of determination of the parameters $\mu$, $\Gamma$ and $\Delta$. However, some more qualitative remarks are in order.

The experiment was conducted at temperature of $5$K and therefore the zero--temperature conductivity considered above is sufficient to describe the observed data. The dependence of best fit value of $\G$ on the magnetic filed was found to be about $\pm10$\%, and we were not able to resolve it further more at the given accuracy of the experimental data. Our calculations showed almost no dependence on the mass gap $\Delta$, provided it is much smaller the chemical potential $\mu$. On the other hand, a mass gap close to the value of chemical potential does change the curves significantly, as on Fig. \ref{fit_alt}. However, one should not consider such fitting procedure as a strong indication of a mass--generation in given graphene samples. It rather calls for more detailed investigation of the phenomena at low fields. The latest research, \cite{Crassee2012}, proves that the spectral shape of the Faraday rotation and the transmission can be affected by magneto--plasmonic effects which are of greater importance for lower fields.
These effects occur due to nanoscale inomogeneities in epitaxial graphene which, in principle, can be taken
account by introducing a background field in the Dirac action. This is an interesting problem, that we
are going to address in the future.

In the original paper \cite{giant_F}, the experiment was fitted using a linear order approximation in conductivity, the later described by the Drude model. It was found there that the cyclotron frequency dependence on magnetic field does not
conform the theoretical prediction (\ref{w0}), $\o_c\approx L_1/2\mu$, first obtained in \cite{Gusynin_NJP11_09}.

Apart from the deviation of the cyclotron frequency from its' theoretical value
\cite{giant_F}, the Drude model was found to be nicely describing the experiment. Still the Drude-like behavior of the Dirac model cannot be uniformly extended for higher frequencies since the later does not predict any poles of conductivity in the complex plane showing a much more complicated behavior.

Finally, we would like to mention here that calculations performed in the framework of the equation of motion method \cite{EqOM} are also in a good agreement with the experimental results of \cite{giant_F}.

\begin{figure}
\centering \includegraphics[width=8cm]{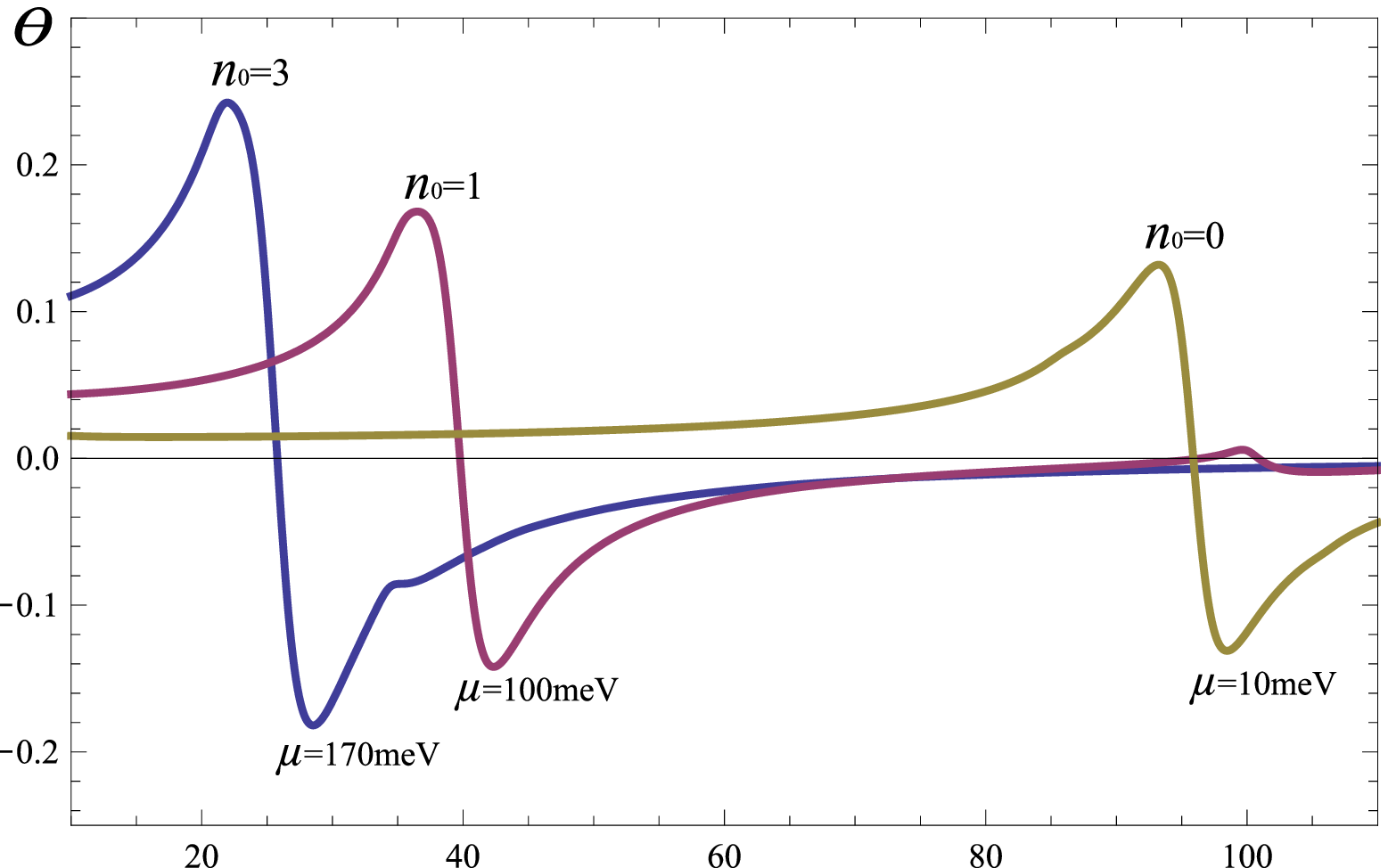}
\centering \includegraphics[width=8cm]{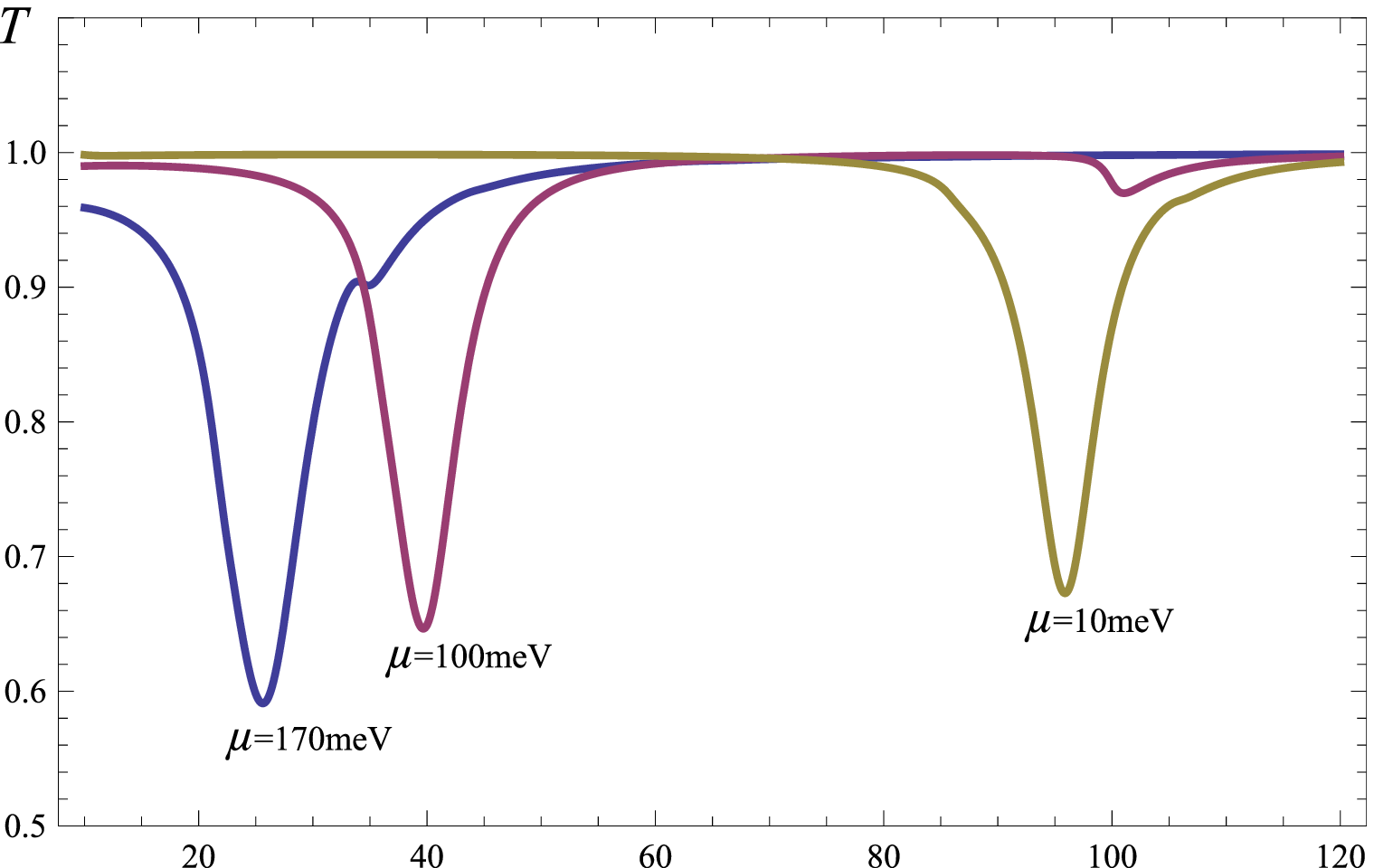}
\caption{%
$\theta$ and $T$ as a function of frequency $p_0$[meV] for different chemical potentials for ${\bf B}=7$T, $\G=1$meV, $\Delta=0$.}
\label{on_p0}
\end{figure}

\subsection{Suspended graphene: moderate and small chemical potential}
\label{small_mu}

We continue by investigating the Faraday rotation as a function of frequency for regimes different of that of \cite{giant_F} --- for moderate and small chemical potentials. The former case might be applicable to the description of an epitaxial graphene subject to the gate voltage, while the later one is relevant for suspended graphene samples or for decoupled graphene layers~\cite{pureGr}.

Analyzing the conductivities similarly to the case of large $\mu$, one can show that for small or moderate chemical potential the biggest contribution to the sum  (\ref{T_th_wrk}) in the vicinity of (\ref{poles}) comes from the $n_0$--term.
However,  in this case the extremum of the real part of $G_{n_0}/p_0$ lie at $p_0\sim\(M_{n_0+1}-M_{n_0}\)\pm \G$, where
\be
    {\rm Re} G_{n_0}\simeq \pm \frac{0.06 }{\G}\(1-\frac{\Delta^2}{M_{n_0+1} M_{n_0}}\),
\ee
here we assumed that $\G\ll M_{n_0+1}-M_{n_0}$. For a small mass gap, $\Delta$, this value is almost independent of the Landau level number, and thus the maximum of rotation angle is inversely proportional to the difference between corresponding Landau levels
\be
    \theta_{max}(\Delta=0) \sim \frac{0.24 f_s \alpha }{\Gamma}\frac{L_b\(1+\delta_{0 n_0}\)}{\sqrt{n_0+1}- \sqrt{n_0}}. 
    \label{th_max}
\ee
For $\Delta=0$ also $M_0=0$, so that two points, $p_0=M_1+M_0$ and $p_0=M_1-M_0$ coincide. Consequently,
one has two identical contributions for $n_0=0$, which is taken into account by $\delta_{0n_0}$
in the formula above.

The expression (\ref{th_max}) gives a rough order of magnitude estimate of the Faraday rotation at the $\a^1$ level. For describing the effect in very clean graphene samples (such as, e.g., reported in \cite{pureGr} where $\G\sim3\cdot10^{-2}$meV) one has to use  complete expression (\ref{Tth}) to correctly take into account the $O(\a/\G)$ terms.

For a suspended graphene we use (\ref{Tth}) to plot on Fig.~\ref{on_p0} both the Faraday rotation and transmission spectra against the frequency at $\textbf{B}=7$T and $\G=1$meV for three different values of chemical potential. The maximum of rotation angle for such parameters has the order of magnitude predicted by (\ref{th_max}), however, the minima in transmission spectra calculated via (\ref{Tth}) are at least two times shallower than the ones predicted by (\ref{Tth_a}). Still, both effects in this case are very large for an one--atom thick material, and presumably can be measured after minor modifications of the existing experimental set-ups. On the other hand, a very promising experiment was recently proposed in \cite{EqOM}. It was predicted there that with graphene placed in an optical cavity the Faraday rotation can reach huge values in the infrared region (up to 60 degrees) and some modest values in the visible one.


\begin{figure}
\centering \includegraphics[width=8cm]{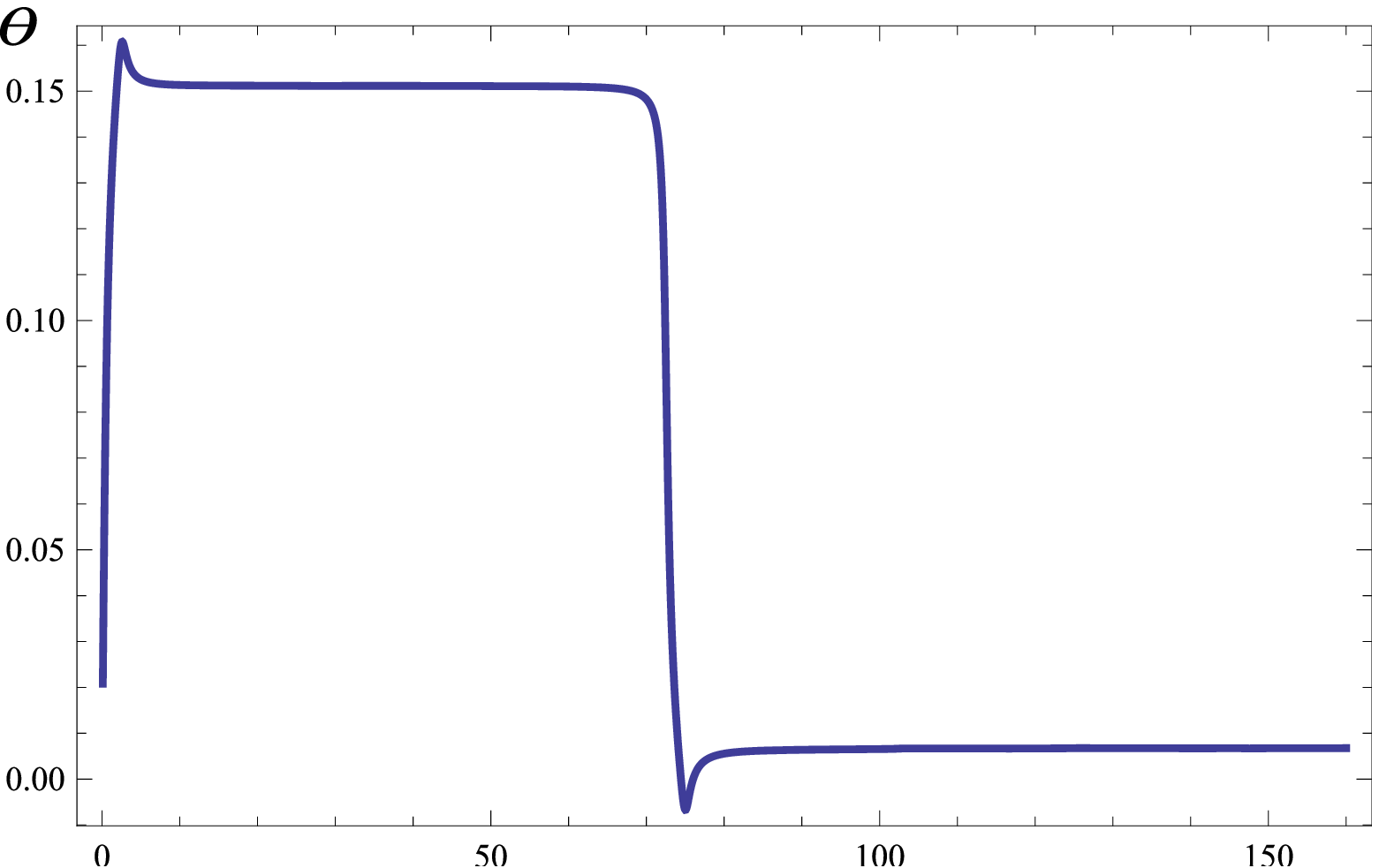}
\includegraphics[width=8cm]{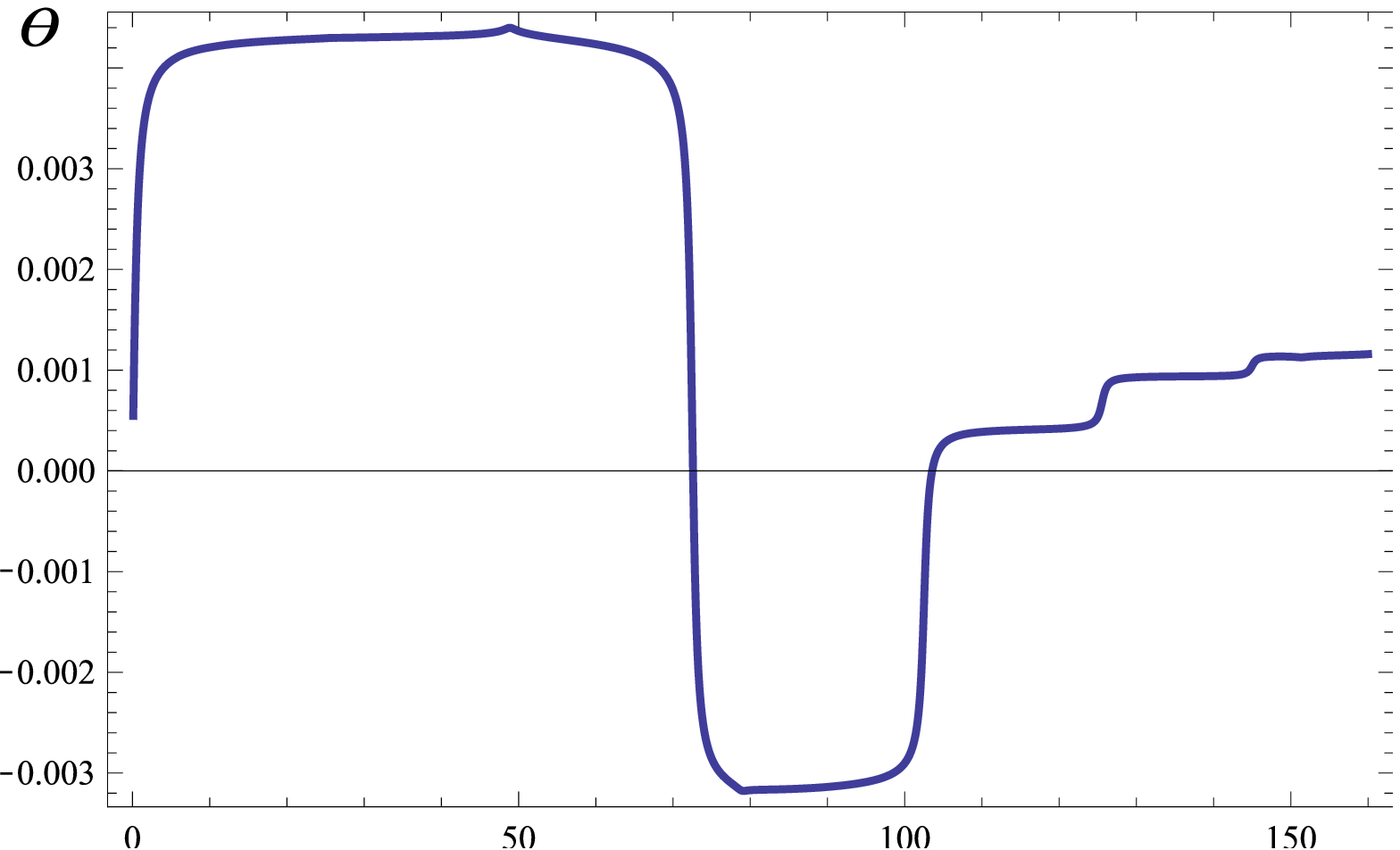}
\caption{
$\theta$ as a function of chemical potential $\mu$[meV], for ${\bf B}=4$T, $\G=0.5$meV for $p_0=75$meV (up), and $p_0=151$meV (down).
\label{on_mu}}
\end{figure}

Finally, we can consider the rotation angle (\ref{Tth_a}) as a function of chemical potential for a fixed frequency.
One would naturally expect that for a small $\G$ its behavior should be similar to the one in the ideal case.
Indeed, a numerical analysis shows that it is quantized in the same manner as the Hall conductivity as defined by (\ref{s_xy_0}).
On the upper picture of Fig. \ref{on_mu}  we plot the Faraday rotation fixing the frequency close to $M_1$. Thus for $\mu<M_1$ the effect gets enhanced according to (\ref{th_max}). On the lower picture the frequency is far away from any special value, and thus the overall amplitude is much smaller.

\section{Conclusions}

In the present paper we recalculated the polarization operator of the Dirac quasi--particles in graphene in external magnetic field and established a clear connection between its components and optical properties of both suspended and epitaxial graphene.

We showed, that in a number of different regimes the rotation of polarization of light passing through monolayer graphene samples is giant being of the order of $0.1$ rad, while the absorption can reach $40$\%.

In particular we showed that the data of the recent experiment on Faraday rotation  in the cyclotron resonance regime  can be nicely fitted by predictions of the Dirac model for sufficiently high magnetic fields (starting with $2$T). Moreover, we predict that for smaller chemical potentials (characteristic for suspended graphene) the effect is still very pronounced. We envisage the pole--like peaks in the polarization rotation angle and the transmission spectra at relatively high frequencies. We also reveal that in clean graphene samples the Faraday rotation as a function of chemical potential should be quantized in a similar fashion as the unconventional Integer Hall effect in graphene.

\section*{Acknowledgements}
We are grateful to A. Kuzmenko
for providing us with the experimental data and the notes \cite{pvt}, as well as for for valuable discussions. The correspondence with M.~Potemski is gratefully acknowledged.

This work was supported in parts by FAPESP (I.V.F. and D.V.V.) and by CNPq (D.V.V.).
\appendix

\end{document}